\documentclass[12pt]{article}
\usepackage{epsfig}

\def\beq{\begin{equation}}
\def\eeq#1{\label{#1}\end{equation}}
\def\eeqn{\end{equation}}

\def\beqa{\begin{eqnarray}}
\def\eeqa#1{\label{#1}\end{eqnarray}}
\def\eeqan{\end{eqnarray}}

\let\bar=\overbar

\def\Dslash{\not{\hbox{\kern-4pt $D$}}}
\def\dslash{\not{\hbox{\kern-2pt $\del$}}}

\def\msb{{\bar{\ssstyle M \kern -1pt S}}}

\def\Title#1{\begin{center} {\Large {\bf #1} } \end{center}}

\begin{document}

\Title{B and D physics from the Tevatron}

\bigskip\bigskip

%+\addtocontents{toc}{{\it D. Reggiano}}
%+\label{ReggianoStart}

\begin{raggedright}  

{\it Paola Squillacioti (for the CDF and D0 Collaborations)\\
INFN Pisa and Siena University, ITALY.}
\bigskip\bigskip
\end{raggedright}

The CDF and D0 experiments at the Tevatron $p\bar{p}$ collider established that extensive and detailed exploration of the b-quark dynamics is possible in hadron collisions, with results competitive and supplementary to B-factories. In this paper we review the current state of Tevatron's heavy flavor measurements considering two main categories: searches for non standard model physics (results on rare decays and CP-violation) and determinations of standard model parameters (annihilation in $B \to h^+ h^-$ decays and   $\gamma$ angle measurement through $B\to DK$ modes).

\section{Searches for non-SM physics}

\subsection{Rare $B\to \mu\mu$ decays}
Decays mediated by flavor changing neutral currents such as $B^0_{(s)}\to \mu^+\mu^-$ are highly suppressed in the Standard Model (SM) because they occur only through higher order loop diagrams. The SM expectations for the branching fractions are $(3.24\pm 0.19)\times 10^{-9}$ for $B_s \to \mu^+\mu^-$ and $(1.04 \pm 0.10)\times 10^{-10}$ for $B_d\to \mu^+\mu^-$ \cite{Buras}.
New physics models like MSSM and SUSY boost the branching fractions up to 100 times the SM expectations, so either observation or null results provide crucial information. 
\\The latest CDF measurement uses 3.7 fb$^{-1}$ of data obtaining the upper limits $BR(B_s \to \mu^+\mu^-) < 4.3\times 10^{-8}$ @ 95\% CL and $BR(B_d \to \mu^+\mu^-) < 7.6\times 10^{-9}$ @ 95\% CL.
D0 instead used 6 fb$^{-1}$ of data measuring  $BR(B_s \to \mu^+\mu^-) < 5.1\times 10^{-8}$ @ 95\% CL. Those upper limits are about an order of magnitude above the SM prediction.
\\Shortly after the conference CDF released an updated and improved $B\to \mu\mu$ analysis, which finds an indication of 
a signal, yielding the first two-sided bound on the branching ratio \cite{art_bmumu}.

\subsection{$B^0_s$ mixing phase}
\begin{figure}[!h]
 \begin{center}
\epsfig{file=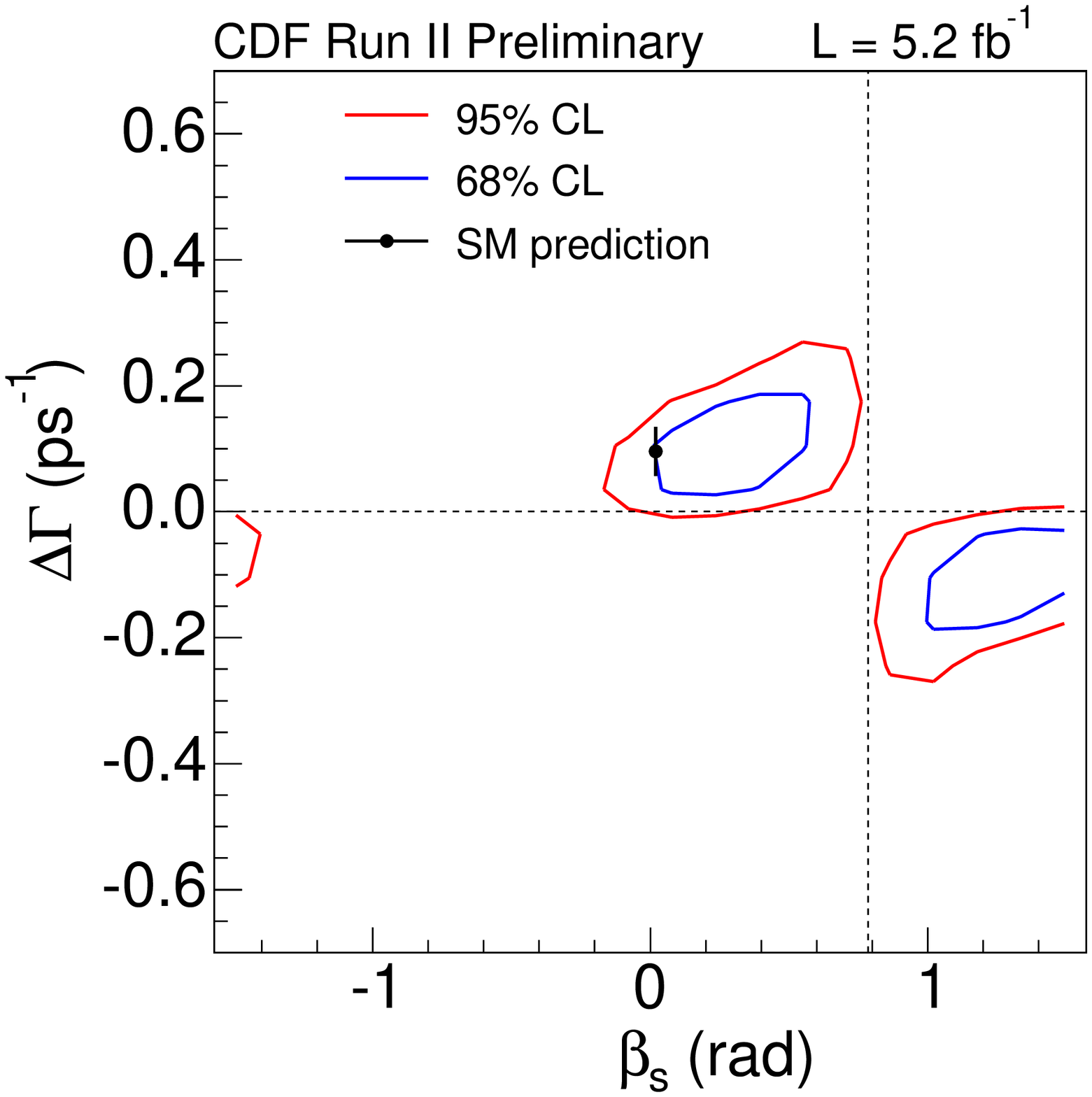,height=1.9in}
\epsfig{file=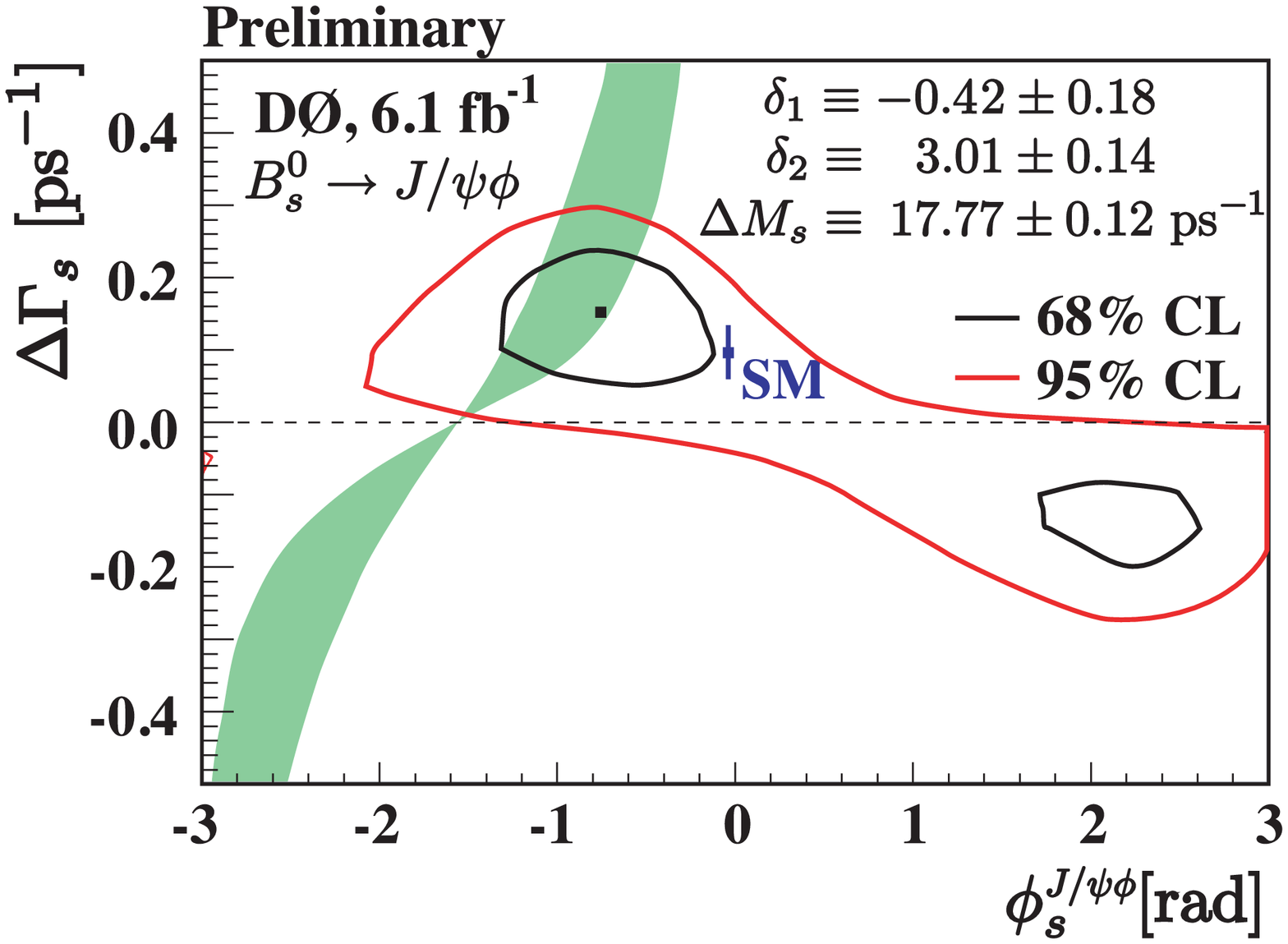,height=1.9in}
\caption{Confidence regions in the $\beta_s^{J/\psi \phi} - \Delta\Gamma$ plane from CDF (left); in the $\phi_{s}-\Delta\Gamma$ plane from D0 (right).}
\label{fig:betas}
\end{center}
\end{figure}

The study of $B_s^0\to J/\psi \phi$ decays allows searching for CP violation effect beyond the SM. In these decays CP violation occurs through the interference between the decay amplitudes with and without mixing. In the SM the relative phase between these decay amplitudes is $\beta_s^{SM}=\arg(-V_{ts}V_{tb}^*/V_{cs}V_{cb}^*)$, expected to be very small. New Physics (NP) contributions in the $B_s^0$ mixing amplitude may alter this mixing phase by a quantity $\phi_s^{NP}$ leading to an observed mixing phase $2\beta_s^{J/\psi \phi}  \simeq 2\beta_s^{SM} - \phi_s^{NP}$. Hence, large values of the observed $\beta_s^{J/\psi \phi}$ would be an indication of physics beyond the SM. 
Confidence regions in the $\beta_s^{J/\psi \phi} - \Delta\Gamma$ plane are constructed by CDF while similar confidence regions are evaluated by D0 in the $\phi_s - \Delta\Gamma$ plane ($2\beta_s^{J/\psi \phi}\approx\phi_s$) (see Fig. \ref{fig:betas}). The mild discrepancy observed in 2008 is now reduced to $1\sigma$.

Both experiments are exploring new ways to probe $\beta_s$, for example the $B_s\to J/\psi f_0(980)$ decay that can be used to measure $\beta_s$ without need of angular analysis. The two experiments measured the $B_s\to J/\psi f_0(980)$ branching fraction with respect to the 
$B_s^0\to J/\psi \phi$ one using similar technique, such as a neural network selection and a simultaneous log likelihood fit for signal and normalization mode.
CDF obtained $R_{f_0/\phi}=\frac{B(B_s\to J/\psi f_0(980))B(f_0(980)\to\pi^+\pi^-)}{B(B_s^0\to J/\psi \phi)B(\phi\to K^+ K^-)}=0.257\pm0.020\pm0.014$ using 3.8 fb$^{-1}$ of data \cite{f0_cdf} and D0 measured $R_{f_0/\phi}=0.210\pm 0.032\pm 0.036$ using 8 fb$^{-1}$ \cite{f0_D0}. These two results are in agreement with Belle and LHCb results \cite{f0_belle_lhcb}.

\subsection{Di-muon charge asymmetry}
In the SM the flavor specific asymmetry in semileptonic b-hadron decays is predicted to be small. It can be generated by an asymmetry in the mixing rate between $b$ and $\bar{b}$ mesons. D0 measured this asymmetry using 6 fb$^{-1}$ of data obtaining $A^b_{sl}=(-0.957\pm0.251\pm0.146)\%$ that is significantly different from the SM prediction $A^b_{sl}=(-0.023^{+0.005}_{-0.006})\%$ \cite{dimuon_theor}. Note from Fig. \ref{fig:betas}, where the green band is obtained from $A_{sl}$, that results from $B_s^0\to J/\psi \phi$ are consistent with the di-muon charge asymmetry. This result has gathered much attention from theorists, and deserves independent confirmation. CDF plans to pursue this measurement with a different technique, in which precise impact parameter information is associated to muons so that they can be unambiguously ascribed to heavy flavor decays.
\\As a fist step, CDF measured the time integrated mixing probability $\bar{\chi}_b$, that measures the probability for a B hadron to mix, using a subset of the same data that will be used for a measurement of $A_{sl}$ and is therefore seen as a preparatory exercise for the $A_{sl}$ analysis. The measurement uses 1.5 fb$^{-1}$ of data collected by CDF and the fundamental ingredient is the use of the muon impact parameter to identify the source of muons. The time integrated mixing is determined to be $\bar{\chi}_b=0.126 \pm 0.08$, in good agreement with the world average.

\subsection{First search of CPV in $B_s\to \phi\phi$ decays}
The $B_s\to \phi\phi$ decay belongs to the class of transitions of pseudoscalar mesons into two vector particles (P $\to$ VV), whose rich dynamics involves three different amplitudes corresponding to the polarization states. 
The $B_s\to \phi\phi$ decay is sensitive to the CP-violating in the interference between decay with and without mixing. Actually, the CP-violating weak phase is predicted to be extremely small in the SM and measurement of nonzero CP-violating observables would indicate unambiguously NP.
Present statistics of the $B_s\to \phi\phi$ data sample are not sufficient for a suitable time-dependent analysis of mixing induced CP-violation as the case of the $B_s\to J/\psi \phi$. However, an investigation of genuine CP-violation observables which could reveal the presence of NP, such as triple products (TP) correlation, is accessible \cite{ref_tp}.
There are two triple products ($u$ and $v$) in the $B_s\to \phi\phi$ decay that are function of the helicity angles.
The CDF collaboration has made the first measurement of $A_u$ and $A_v$ asymmetries in $B_s\to \phi\phi$ using 2.9 fb$^{-1}$ of data \cite{ref_Bsphiphi}. The asymmetries are obtained through an unbinned maximum likelihood fit. The background asymmetries are consistent with zero and the final results for signal asymmetries are $A_u = (-0.7\pm6.4(stat.)\pm(syst.))\%$ and $A_v = (-12.0 \pm 6.4(stat.) \pm 1.6(syst.))\%$.

\subsection{CPV in $D \to h^+ h^-$}
CDF has today the world's largest charm samples. This offers the opportunity to pursue a rich analysis program that includes access to direct CP violating asymmetries, branching fractions, mixing and mixing-induced CP violation. Examples of clean channels with possible additional sources of CP violation in the charm system are $D^0\rightarrow \pi^-\pi^+$ and $D^0 \rightarrow K^- K^+$. Contribution to these decays from ``penguin'' amplitudes are negligible in the SM, so the presence of NP particles could enhance the size of CP violation with respect to the SM expectation. Any asymmetry significantly larger than a few times 0.1\% may ambiguously indicate NP contribution.
\\CDF measures the asymmetry using $D^0\rightarrow h^-h^+$ decays from charged $D^*$ mesons through fits of the $D^0\pi$ mass distributions.
In 6 fb$^{-1}$ of data CDF reconstructs a huge sample of $D^* \rightarrow D^0\pi$ candidates: 215000 $D^*$-tagged $D^0\rightarrow \pi^-\pi^+$ (and charge conjugate) decays, 476000 $D^*$-tagged $D^0\rightarrow K^-K^+$ (and c. c.) decays, 5 million $D^*$-tagged $D^0\rightarrow K^-\pi^+$ (and c. c.) decays and 29 million $D^0\rightarrow K^-\pi^+$ (and c. c.) decays where no tag was required. 
\\The final results are $A_{CP}(D^0\rightarrow \pi^-\pi^+) = [+0.22 \pm 0.24 (stat.)\pm 0.11(syst.)]$ and $A_{CP}(D^0\rightarrow K^-K^+) =  [-0.24 \pm 0.22 (stat.)\pm 0.10(syst.)]$ which are consistent with CP conservation and also with SM predictions, and are th most stringent to date.

\section{Determination of SM parameters}
\subsection{First evidence of annihilation in $B \to h^+ h^-$ decays}
Two-body non-leptonic charmless decays of b-hadrons are largely studied processes in flavor physics. Some decays receive contributions from higher-order (``penguin'') transitions, and are therefore sensitive to the possible presence of NP in internal loops. 
The $B_s^0 \to \pi^+\pi^-$ and $B^0 \to K^+K^-$ decay modes have a special status in that all quarks in the final state are different from those in the initial state. This limits the possible diagrams that contribute to these decay to penguin-annihilation (PA) and W-exchange (E) topologies. These amplitudes are difficult to predict within the current phenomenological models. and are often neglected in calculations. Estimates of these amplitudes in the QCD factorization (QCDF) approach \cite{qcdf} are affected by significant uncertainties. Recent perturbative QCD calculation (pQCD) provide some potentially testable predictions \cite{pqcd}.

CDF reports the results of a simultaneous search for the two decays $B_s^0 \to \pi^+\pi^-$ and $B^0 \to K^+K^-$ using 6 fb$^{-1}$ of data. Fig. \ref{fig:bhh} shows the $B \to h^+ h^-$ invariant mass distribution obtained assigning the charged pion mass to both decay products. 
An extended unbinned likelihood fit, incorporating kinematic (kin) and particle identification (PID) information provided by the specific ionization ($dE/dx$) in the CDF drift chamber, is performed to determine the fraction of each individual mode in our sample.  
CDF obtains $94\pm 28 \pm 11$ $B_s^0 \to \pi^+\pi^-$ and $120 \pm 49 \pm 42$ $B^0 \to K^+K^-$ signal events. CDF observes the first evidence of  $B_s^0 \to \pi^+\pi^-$ with a significance of 3.7 $\sigma$, while no evidence is found for $B^0 \to K^+K^-$ mode (2 $\sigma$ of significance).

CDF quotes relative and absolute branching fractions for the two modes. The branching fraction measured for the $B_s^0 \to \pi^+\pi^-$ mode ($(0.57 \pm 0.15 \pm 0.10)\times 10^{-6}$) agrees with predictions obtained with the pQCD approach \cite{pqcd}, but it is higher than most other theoretical predictions \cite{qcdf}.
The measurement of the branching fraction for $B^0 \to K^+K^-$ ($(0.23 \pm 0.10 \pm 0.10)\times 10^{-6}$) is the world's best and it is in agreement with other existing measurements  \cite{bkk_prev} and with theoretical predictions \cite{qcdf}.

\subsection{$\gamma$ angle from $B\to DK$ modes}
The measurement of CP-violating asymmetries and branching ratios of $B \to DK$ modes allows a theoretically-clean extraction of the CKM angle $\gamma$.  Using these decays $\gamma$ could be extracted by exploiting the interference between the tree amplitudes of the $b\rightarrow c\bar{u}s$ ($B^- \rightarrow {D}^0 K^-$) and $b\rightarrow u\bar{c}s$ ($B^- \rightarrow \overline{D}^0 K^-$ ) processes. This can be obtained in several ways, using different choices of $D$ decay channels.

A class of decays that has been theoretically studied are $B^-\to DK^-$ decays that are a coherent superposition 
of the color favored $B^- \rightarrow D^0 K^-$ followed by the doubly Cabibbo suppressed decay $D^0\to K^+\pi^-$, and of the color suppressed $B^- \rightarrow \bar{D}^0 K^-$ followed by the Cabibbo favored decay $\bar{D}^0 \to K^+ \pi^-$.

The following observables can be defined \cite{ADS}:
 
\begin{eqnarray}
\displaystyle R_{ADS} & = & \frac{\mathcal{B}(B^-\rightarrow [K^+ \pi^-]_{D}K^-)+\mathcal{B}(B^+\rightarrow [K^-\pi^+]_{D}K^+)}{\mathcal{B}(B^-\rightarrow [K^- \pi^+]_{D}K^-)+\mathcal{B}(B^+\rightarrow [K^+\pi^-]_{D}K^+)}\\
\displaystyle A_{ADS} & = & \frac{\mathcal{B}(B^-\rightarrow [K^+\pi^-]_{D}K^-)-\mathcal{B}(B^+\rightarrow [K^-\pi^+]_{D}K^+)}{\mathcal{B}(B^-\rightarrow [K^+\pi^-]_{D}K^-)+\mathcal{B}(B^+\rightarrow [K^-\pi^+]_{D}K^+)}.
\end{eqnarray}
where $B^-\rightarrow [K^+ \pi^-]_{D}K^-$ is the suppressed ($sup$) mode and $B^-\rightarrow [K^- \pi^+]_{D}K^-$ is the favored ($fav$) mode.
These quantities are related to the CKM angle $\gamma$ by the equations $R_{ADS}  =  r_D^2 + r_B^2 + 2r_Dr_B \cos{\gamma}\cos{(\delta_B+\delta_D)}$ and $A_{ADS}  =  2r_Br_D\sin{\gamma}\sin{(\delta_B+\delta_D)}/R_{ADS}$,
where $r_B$ is the magnitude of the ratio of the amplitudes of the processes $B^-\rightarrow \overline{D}^0 K^-$ and $B^- \rightarrow D^0 K^-$, and $\delta_B$ is their relative strong phase; $r_D$ is the magnitude of the ratio of the amplitudes of the processes $D^0\rightarrow K^-\pi^+$ and $D^0 \rightarrow K^+\pi^-$, and $\delta_D$ is their relative strong phase. 

\begin{figure}[ht]
\begin{minipage}[b]{0.4\linewidth}
\centering
\includegraphics[scale=0.22]{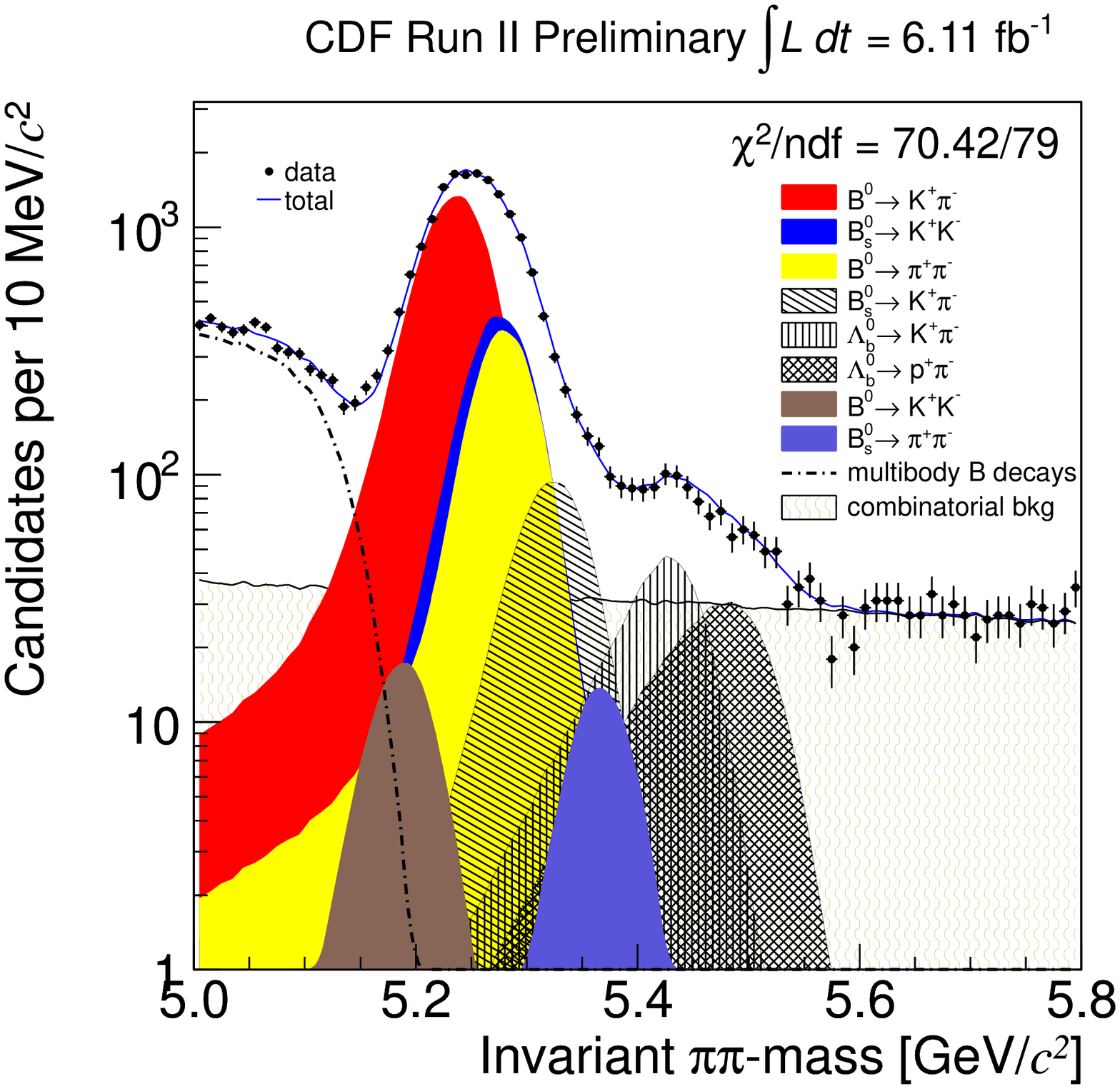}
\caption{$B \to h^+ h^-$ mass distribution. The charged pion mass is assigned to both tracks. The projections of the likelihood fit are overlaid.}
\label{fig:bhh}
\end{minipage}
\hspace{0.5cm}
\begin{minipage}[b]{0.6\linewidth}
\centering
\includegraphics[scale=0.22]{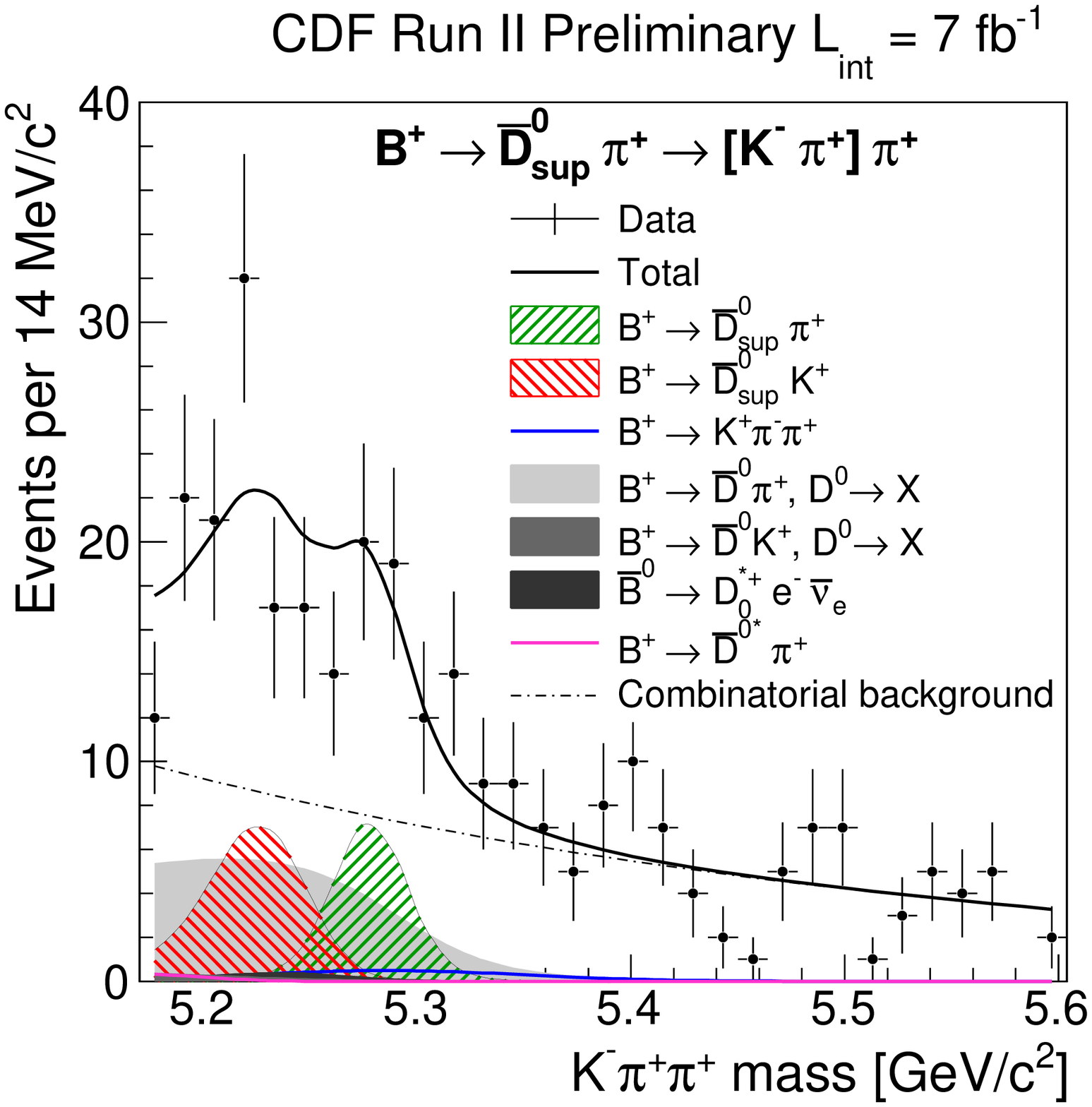}
\includegraphics[scale=0.22]{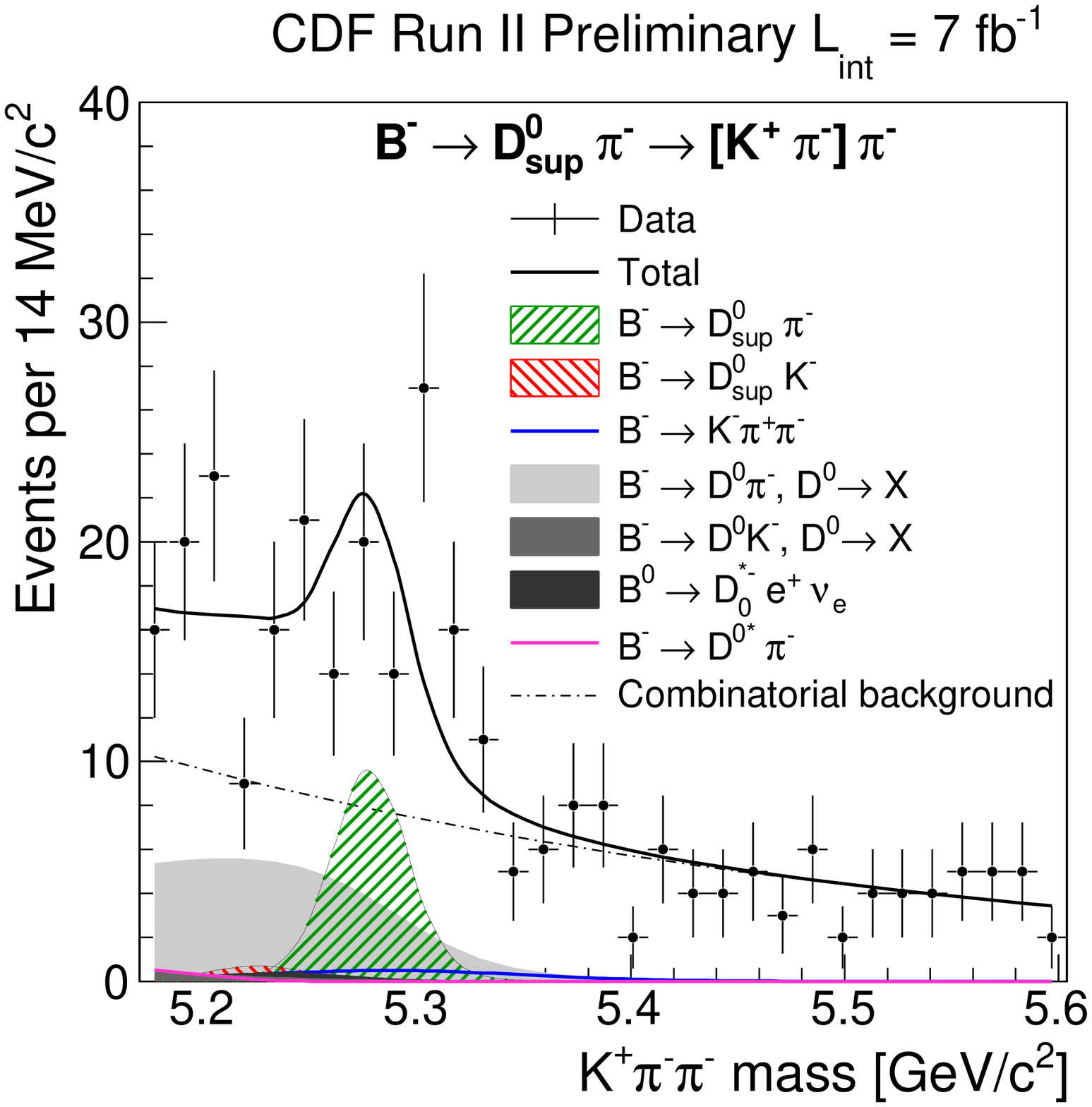}
\caption{Left: Mass distributions of $B^{+} \rightarrow \bar{D}_{sup} h^{+}$
candidates. Right: Mass distributions of $B^{-} \rightarrow D_{sup} h^{-}$
candidates. The pion mass is assigned to the track from the $B$ decay. The projections of the likelihood fit are overlaid.}
\label{fig:ads_proj}
\end{minipage}
\end{figure}

CDF describes the first reconstruction of  $B^-\to D_{\it sup}K^-$ modes performed in hadron collisions based on a total integrated luminosity of 7 fb$^{-1}$ of data \cite{dcs_paper}.
An unbinned likelihood fit, exploiting mass and particle identification information provided by the 
specific ionization ($dE/dx$) in the drift chamber, is performed to separate the $B \rightarrow DK$ contributions from the $B \rightarrow D\pi$ signals, from the combinatorial background and from the physics backgrounds.
Fig. \ref{fig:ads_proj} shows the suppressed mode invariant mass distributions separated in charge. CDF obtaines $32\pm 12$ $B \to D_{sup} K$ and $55\pm 14$ $B \to D_{sup} \pi$ signal events. CDF observes the first evidence of  $B \to D_{sup} K$ signal at a hadron collider with a significance of 3.2 $\sigma$.
CDF measures the asymmetries $A_{ADS} (K)= -0.82 \pm 0.44(\rm{stat}) \pm 0.09 (\rm{syst})$ and $A_{ADS} (\pi)= 0.13 \pm 0.25(\rm{stat}) \pm 0.02 (\rm{syst})$ and the ratios of doubly Cabibbo suppressed mode to flavor eigenstate $R_{ADS} (K)= [22.0 \pm 8.6(\rm{stat}) \pm 2.6 (\rm{syst})]\cdot 10^{-3}$ and $R_{ADS} (\pi)= [2.8 \pm 0.7(\rm{stat}) \pm 0.4 (\rm{syst})]\cdot 10^{-3}$.
The results are in agreement with existing measurements performed at the
$\Upsilon(4S)$ resonance.

%%
%%   use this format to include a LaTeX table  into your paper
%%
%\begin{table}[b]
%\begin{center}
%\begin{tabular}{l|ccc}  
%Patient &  Initial level($\mu$g/cc) &  w. Magnet &  
%w. Magnet and Sound \\ \hline
% Guglielmo B.  &   0.12     &     0.10      &     0.001  \\
% Ferrando di N. &  0.15     &     0.11      &  $< 0.0005$ \\ \hline
%\end{tabular}
%\caption{Blood cyanide levels for the two patients.}
%\label{tab:blood}
%\end{center}
%\end{table}
%%%%%%%%%%%%%%%%%%%%%%%%%%%%%%%%%%%%%%%%%%%%%%%%%%%%%%%%%%%%%%%%%%%%%%%%%%%

\end{document}